\pdfoutput=1

\documentclass[twoside,leqno,twocolumn]{article}
\usepackage{ltexpprt}

\usepackage{graphicx}
\usepackage{dcolumn}

\usepackage{fancybox}
\usepackage{amsmath}
\usepackage{amssymb}

\usepackage{url}
\usepackage{color}

\begin{document}









\title{\Large Redesigning pattern mining algorithms for supercomputers}
\author{Kazuki Yoshizoe
\thanks{Graduate School of Frontier Sciences, The University of Tokyo, Kashiwa, Japan}
\thanks{Tokyo Institute of Technology, Tokyo, Japan}
\\
{\small zoe.f16xl@gmail.com}
\and
Aika Terada
\footnotemark[1]
\thanks{Research Fellow of Japan Society for the Promotion of Science}
\thanks{Biotechnology Research Institute for Drug Discovery, National Institute
of Advanced Industrial Science and Technology,
Tokyo
, Japan}\\
{\small terada@cbms.k.u-tokyo.ac.jp}
\and
Koji Tsuda
\footnotemark[1]
\footnotemark[4]
\thanks{Center for Materials Research by Information Integration,
National Institute for Materials Science,
Tsukuba
, Japan}
\\
{\small tsuda@k.u-tokyo.ac.jp}
}
\date{}

\maketitle


\begin{abstract} 
Upcoming many core processors are expected to employ a distributed
memory architecture similar to currently available supercomputers, 
but parallel pattern mining algorithms amenable to the architecture 
are not comprehensively studied.
We present a novel closed pattern mining algorithm with a well-engineered 
communication protocol, and generalize it to find statistically
significant patterns from personal genome data.
For distributing communication evenly, it employs global load balancing with
multiple stacks distributed on a set of cores 
organized as a hypercube with random edges.
Our algorithm achieved up to 1175-fold speedup by using 1200 cores
for solving a problem with 11,914 items and 697 transactions, 
while the naive approach of separating the search space failed completely.
 
\end{abstract}

\section{Introduction}
Parallel algorithms for pattern mining have been a long-standing subject
of research~\cite{agrawal1996parallel,paraminer,Tatikonda:2009:MTD:1687627.1687706,DBLP:conf/icdm/BuehrerPC06,DBLP:conf/ieeehpcs/NegrevergneTMU10}.
Algorithms for shared memory environments~\cite{paraminer} are losing
ground, because upcoming many-core systems such as Intel
Single-Chip Cloud Computer (SCC)~\cite{corley2010intel} will inevitably employ a distributed
memory architecture due to difficulty in concurrent memory access.

This paper aims to redesign pattern mining algorithms for distributed
memory environments. Parallel search algorithms distribute the search
tree by maintaining yet unexplored nodes in one or multiple stacks, 
and letting a processing core pick up a node and explore the search tree further down.
Newly found nodes are stored back to stacks and they will then be taken by other cores.
In a distributed memory environment, holding the stacks
at one core causes severe concentration of communication.
Thus, distribution of stacks to cores and efficient communication
among them is a central issue that does not arise in shared memory studies~\cite{paraminer}.  
Our algorithm generalizes the LCM algorithm~\cite{uno2004lcm} based on
global load balancing with multiple stacks distributed on a set of cores 
organized as a hypercube with random edges~\cite{Saraswat:2011cd}.
This structure is effective in distributing communication necessary 
for pruning the search tree. 

Our interest in parallelization is motivated by ever growing personal
genome data, i.e., mutation profiles of individuals~\cite{maher2008personal}.
Among several pattern mining-based approaches for genetics studies~\cite{onkamo2006survey}, 
we picked up limitless-arity multiple testing procedure (LAMP)~\cite{Terada2013,Minato2014}, 
because it computes properly corrected P-values for each pattern of
alleles, and allows us to find all statistically significant patterns.
Our strategy performed extremely well in parallelizing LAMP: 
Up to 1175-fold speedup was observed with 1200 cores.

The rest of this paper is organized as follows.
Basics about closed itemset mining and parallel search are briefly
reviewed in Section~\ref{sec:prelim_lcm_parallel}.
LAMP is outlined in Section~\ref{preliminary}.
Section~\ref{sec:method} and~\ref{sec:experiments}
describe our method and experimental results, respectively.
Section~\ref{sec:related_work} describes related work
and Section~\ref{sec:conclusion} concludes our paper.

\section{Backgrounds}
\label{sec:prelim_lcm_parallel}

\subsection{Linear time Closed itemset Miner (LCM)}

\begin{figure}
  \begin{center}
    \includegraphics[scale=0.4]{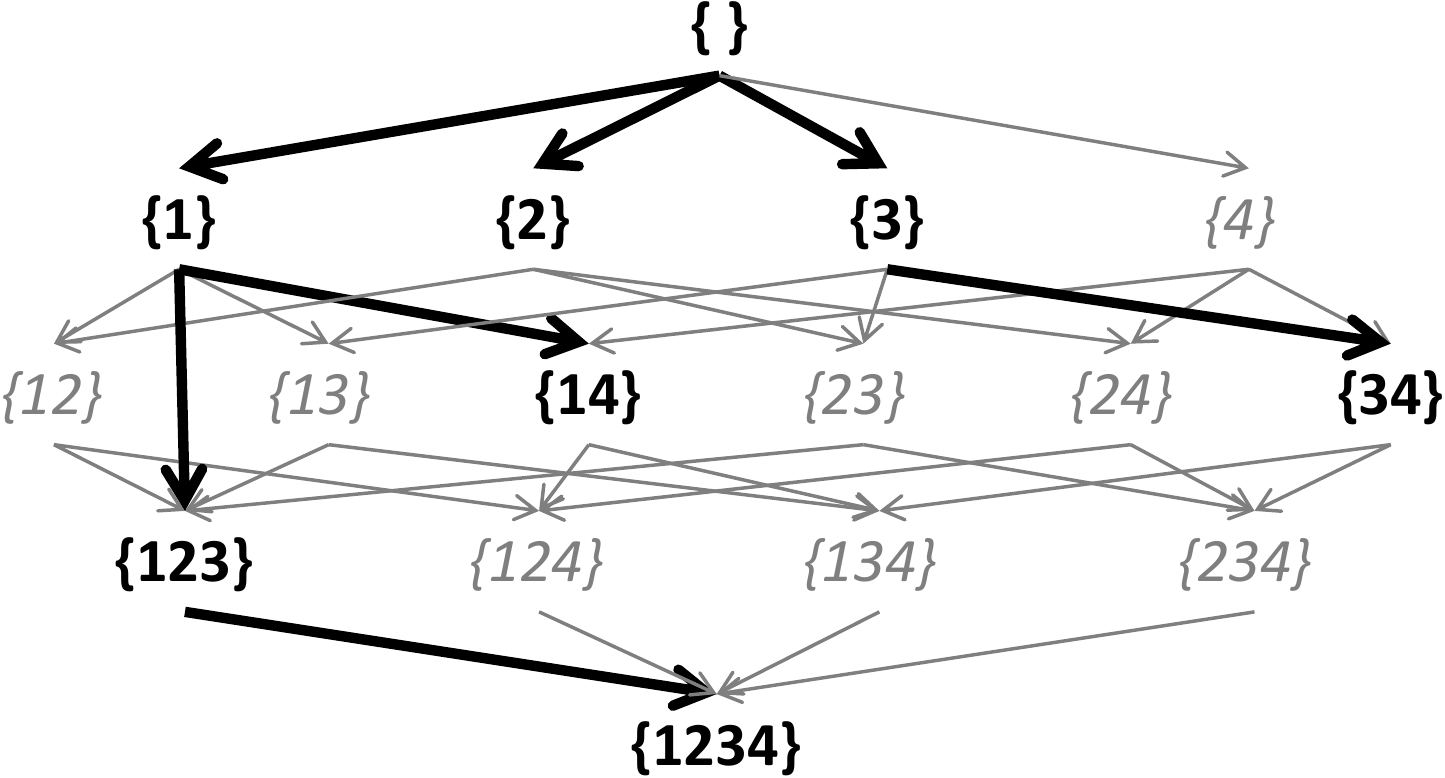}
    \caption{Original search space and LCM tree}
    \label{fig:lcm_tree}
  \end{center}
\end{figure}

Given an itemset $I$
such that there is no item $j \notin I$ which satisfies
$sup(I) = sup(I \cup {j})$,
$I$ is a {\it closed itemset}.
In other words, if an item is added to
a closed itemset, the support will always become smaller.
Closed itemset provides a loss-less compressed
expression for enumerating itemsets.
Original search space of itemset mining is shown in gray
itemsets and edges in Fig.~\ref{fig:lcm_tree}.
Linear time Closed itemset Miner (LCM)
is a famous technique in frequent itemset mining~\cite{uno2004lcm},
which modifies the search space to a tree
with edges connecting only closed itemsets.
Please refer to~\cite{uno2004lcm} for the details of the LCM algorithm.

\subsection{Parallel Search}
\label{sec:prelim_parallel_search}

There is no apparent parallelism in search algorithms because
the subsequent behavior gets affected by the preceding results.
Therefore, search algorithms are one of the difficult targets for parallelization.

One of the naive approaches
is to divide the original tree
to a number of subtrees
and assign the subtrees to processes.
Of course, it does not perform well
if the search tree is unbalanced.
In practical search problems,
trees are normally highly unbalanced because of
pruning techniques.

To equally distribute the workload to processes,
parallel search techniques have been developed for
popular search algorithms (\cite{Saraswat:2011cd},
\cite{Kishimoto:2013:ESS:2435476.2435960},\cite{Romein:1999vy}).
Fortunately, depth-first search on trees
is one of the most feasible problem among parallel search.
It could be parallelized by transforming
the algorithm to queue based version and
using dynamic load balancing (explained in Section~\ref{sec:method}).




\section{Significant Pattern Mining} \label{preliminary}

In this section, 
we explain significant pattern mining introduced by Terada et al.~\cite{Terada2013}. 



Suppose that we have $N$ transactions. 
Each transaction contains a set of items 
and is classified into positive or negative. 
In total, $N_{pos}$ transactions are categorized into positive. 
The goal of significant pattern mining is 
to enumerate statistically significant associations 
between itemsets and the classification such that
the probability that at least one false discovery appears,
which is called the family-wise error rate (FWER), 
is at most $\alpha$. 

\subsection{Statistical assessment for an itemset}

Given an itemset $I$, 
we assess the statistical significance of $I$ as the following. 
Let $x(I)$ and $n(I)$ be the frequency of $I$ 
in all transactions and in positive transactions, respectively. 
The P-value of $I$ is calculated by the one-sided Fisher's exact test as
\begin{eqnarray*}
  P(I) = \sum_{n_i = n(I)}^{\min{ \{ x(I), N_{pos}\}}}
  \frac{
    \left ( \begin{array}{c} N_{pos}\\n_i \end{array} \right )
    \left ( \begin{array}{c} N - N_{pos}\\x(I) - n_i \end{array} \right )}
    {\left ( \begin{array}{c} N\\x(I) \end{array} \right )}.
\label{eq:fisher}
\end{eqnarray*}
On testing for a single itemset, when $P(I) \leq \alpha$, 
$I$ is regarded as significantly associated positives.

On performing statistical assessments on multiple itemsets, 
FWER
becomes large. 
Therefore, 
the multiple testing procedure should be conducted 
to adjust the threshold for the P-value 
so that FWER becomes at most $\alpha$~\cite{Dudoit2007}. 
For example, given $k$ hypotheses, 
the Bonferroni correction, which is a widely used multiple testing procedure, 
calibrates the threshold such that $\delta = \alpha/k$. 
The itemset whose P-value is up to $\delta$ 
is regarded as significantly associated with the classification. 

In itemset mining, 
detection of a significant itemset is hopeless 
using the Bonferroni correction 
because $k$ exponentially increases to the number of items. 


\subsection{LAMP}
\label{subsec:lamp}

More recently, 
the multiple testing procedure, which is named LAMP, 
has been proposed for detecting significant itemsets~\cite{Terada2013}. 
LAMP achieves higher sensitivity than the Bonferroni correction, 
whereas the upper bound of FWER is under the same level. 
We introduce here key points of LAMP: 
(1) Extremely low frequent itemsets can be eliminated from the Bonferroni factor $k$. 
(2) LAMP solves an optimization problem about the threshold for closed itemset frequency.

The first key point is obtained from the Tarone's P-value bound strategy~\cite{Tarone1990}. 
Given the marginal distribution $N$, $N_{pos}$ and $x$, 
the lower bound of P-value is calculated as
\[
  f(x) = \left ( \begin{array}{c} N_{pos}\\x \end{array} \right ) \bigg /
     \left ( \begin{array}{c} N\\x \end{array} \right ).
\]
If $f(x) > \delta$, 
the itemset can never result in significant without calculation of P-value. 
These itemsets never increase the FWER, 
and hence they are eliminated from Bonferroni factor. 
Because $f(x)$ monotonically decreases to $x$, 
the frequencies of eliminated itemsets are less than the optimal threshold $\lambda$. 

From this insight, 
LAMP first calculates the largest $\lambda$ that satisfies the following condition: 
\begin{eqnarray}
  f(\lambda - 1) > \alpha / \mbox{CS}(\lambda), 
  \label{eq:lamp_condition}
\end{eqnarray}
where $\mbox{CS}(\lambda)$ represents the number of closed itemsets. 
Then $\delta = \alpha / \mbox{CS}(\lambda)$ is used as the adjusted significance level, 
which is normally much larger than that of Bonferroni correction. 

\subsection{Support increase algorithm for LAMP}
\label{subsec:support_increase}

LAMP algorithm follows three phases.
First phase finds an appropriate minimum support
and then the second phase performs a normal frequent closed itemset mining
based on it.
In the third phase, statistically significant
itemsets are extracted from the closed itemsets.

\begin{figure*}
  \begin{center}
    \includegraphics[scale=0.5]{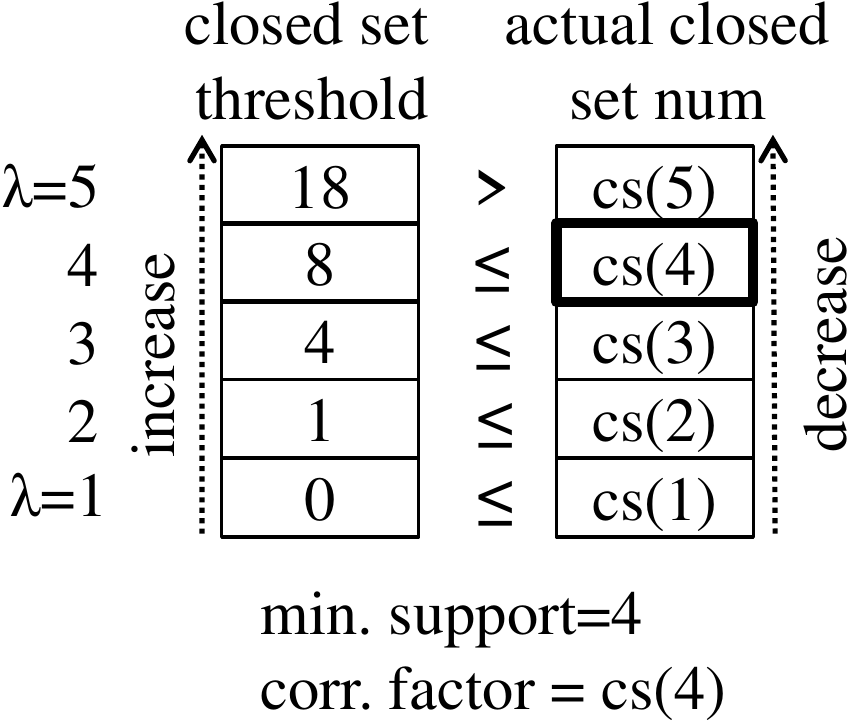}
    \hspace{3mm}
    \includegraphics[scale=0.5]{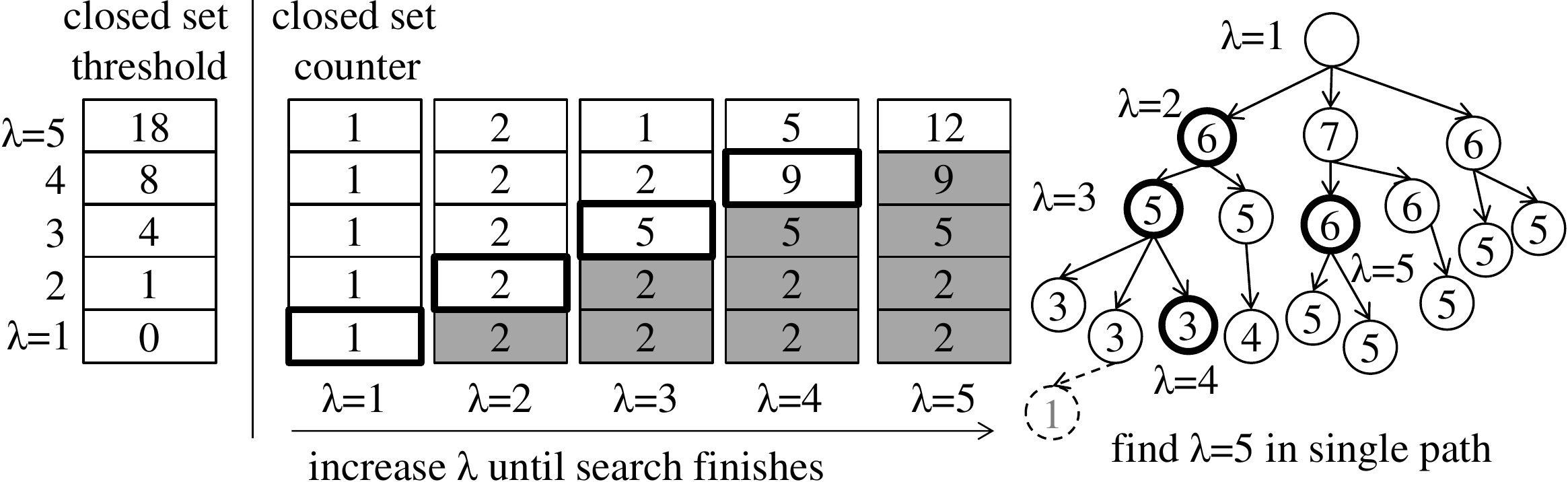}
    \caption{Support increase algorithm}
    \label{fig:support_increase}
  \end{center}
\end{figure*}

The definition of appropriate minimum support comes
from theory as described in Section~\ref{preliminary}
which is illustrated
in the left of Fig.~\ref{fig:support_increase}.
The number of closed itemsets
monotonically decreases as the
$\lambda$ (e.g. frequency) increases.
On the other hand,
from Eq.~\ref{eq:lamp_condition}
the threshold for closed itemset number
can be calculated
which monotonically increases as $\lambda$ increases.
(The numbers in the figure are just examples
but are taken from a small but realistic problem.)

The inequality sign flips at $\lambda=5$.
Subtracting 1 from the $\lambda$ we obtain the
appropriate minimum support, 4 in this case,
and the closed set number $CS(4)$ will be
the correction factor for finding statistically
significant itemsets.

In a simple approach,
it could be found by counting
closed itemset for all possible $\lambda$ value.
However, by using the support increase algorithm~\cite{Minato2014},
minimum support can be found in a single run.
The algorithm is explained in the right of Fig.~\ref{fig:support_increase}.

Initial value of $\lambda$ is set to $0$.
Each node in the tree represents one closed itemset
and the number in the node shows the support.
Please note that the search goes from left child to right child
in depth-first manner.
When the first child is traversed,
the support is $6$, and closed itemsets for $\lambda \leq 6$ becomes 1.
The threshold for $\lambda=1$ is immediately exceeded
and $\lambda$ is incremented. The boxes and nodes with
thick line shows where a threshold was exceeded.
When the next child is traversed, support is $5$.
Again, closed itemsets for $\lambda \leq 5$ are incremented by 1
and the threshold for $\lambda=2$ gets exceeded,
$\lambda$ becomes 3.

This process continues until the search ends
with $\lambda$ value of $5$ without exceeding the threshold for $5$.
The search space shrinks as $\lambda$ increases.
The dotted node with support 1 will be ignored because
when the algorithm reach there, $\lambda$ is already 3.
(More nodes are likely to be
ignored but those are omitted.)
In this way we can efficiently know the {\it minimum support}
which is $4$ in this case.
(It is smaller than the last $\lambda$ by 1.)

Then in the second phase,
the number of closed frequent itemsets with 
minimum support $\lambda$ of 4 is counted
and used as the correction factor.

Lastly, in the third phase,
statistically significant itemsets are chosen from
the discovered closed itemsets.

\section{Proposed Method}
\label{sec:method}

Since our target is a tree,
we will use the term {\it node} to denote
nodes of the tree.
To avoid confusions,
we will use the normal term {\it compute node} only in limited places
and use the term {\it process} instead.
One process will be assigned to one CPU core.
Therefore, there will be 12 processes on a 12 core machine.

\subsection{Parallelize depth first search using stack}

Single thread depth first search can be
simply implemented using recursive function call as
shown in the function DFS in Fig. \ref{alg:DFS}.
If the search reaches the leaf node, there will be no children
and the function returns.
Back-tracking can be implemented naturally in this way.

\begin{figure}
  \begin{center}
    \fbox{ \footnotesize
      \begin{minipage}[t]{.4\textwidth}
        \begin{tabbing}
          DFS(node $n$) \{\\
          \quad\=\textbf{foreach} (child $c$ of $n$) DFS($c$)\\
          \}\\
          DFS\_Loop() \{\\
          \>\textbf{while} (stack not empty)\\
          \>node $n$=Pop()\\
          \>\textbf{foreach} (child $c$ of $n$) Push($c$)\\
          \}
        \end{tabbing}
      \end{minipage}
    }
  \end{center}
  \caption{Depth First Search with recursion and stack}
  \label{alg:DFS}
\end{figure}

Now we describe how to transform this into
a stack based version as a preparation for parallelization.
As shown in the function DFS\_Loop in Fig. \ref{alg:DFS},
we can replace the recursive function call for each child
with push to and pop from the stack.

Only the root node will be pushed to the stack before the
algorithm starts.
Then the top entry of the stack will be popped and
the children of the node will be pushed to the stack.
The algorithm terminates when the stack becomes empty.
For this approach, each node on the stack
must have enough data for search.
For itemset mining, the itemset data itself
identify the node of the search tree.

\begin{figure}
  \begin{center}
    \includegraphics[scale=0.4]{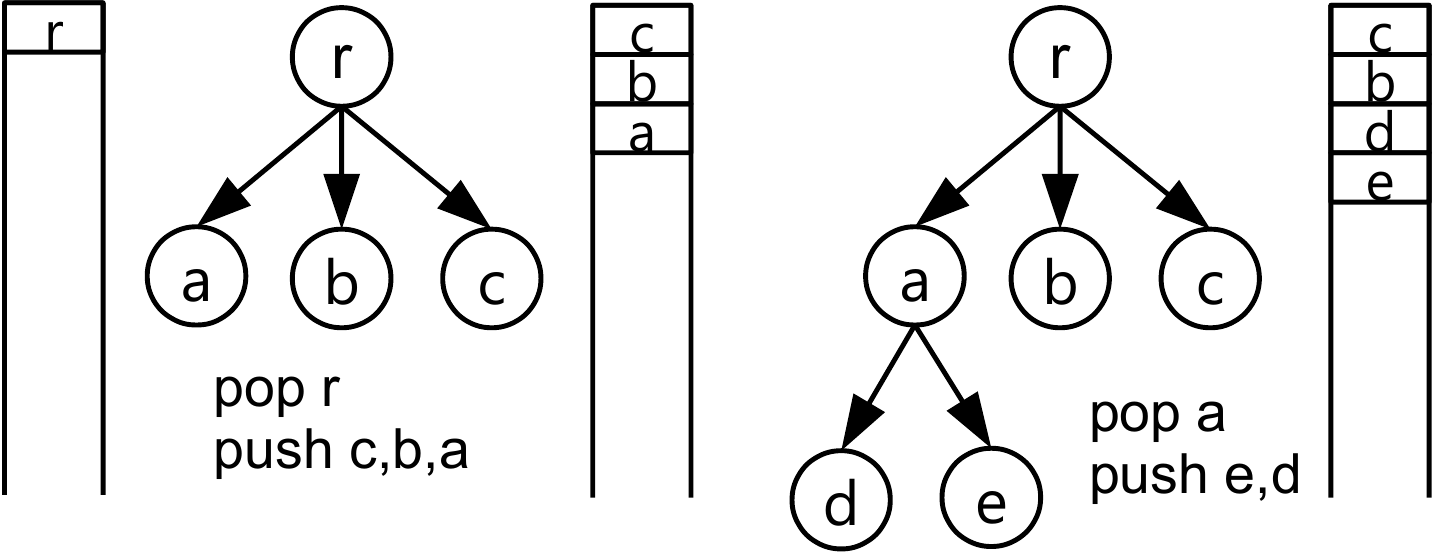}
    \caption{Stack operation in DFS}
    \label{fig:DFS_tree}
  \end{center}
\end{figure}

Fig. \ref{fig:DFS_tree} shows the behavior of DFS\_Loop.
Here we intentionally reversed the order of the children
when pushing nodes onto the stack to make the search order
equivalent with the original DFS function.

The algorithm does work if we change the stack (First-in-last-out)
to a queue (First-in-first-out).
However, if a queue is used, the order of searching nodes
will be equivalent to breadth-first search and,
in general, breadth first search requires more memory because
it stores a large portion of the tree.
With stack (with reverse order push) the search order
will be equivalent to depth-first search
and the stack need to store only the nodes and the sibling nodes
in the current search path.
The required memory size will be limited because it is proportional to
(search depth) $\times$ (number of branch).
The usage is greater than the original DFS by a factor of
(number of branches), but still small enough in practical cases.

\subsection{Parallel search based on dynamic load balancing}

Once we have a search algorithm using the stack,
the workload can be balanced by distributing
the stack to processes.

The main part of the parallel algorithm
is handled by the Probe function
which receives messages and
processes the tasks based on the message type.
Probe relies on a function in the MPI library named MPI\_Iprobe
which returns true if a message is received
and returns false otherwise.
However it could be implemented
on multithread environment with little trouble.

\begin{figure}
  \begin{center}
    \fbox{ \footnotesize
      \begin{minipage}[t]{.4\textwidth}
        \begin{tabbing}
          \textbf{int} $l=2$ // length of hypercube\\
          \textbf{int} $z=$ (dimension of hypercube)\\
          \textbf{int} $w=1$ (number of random steal trials)\\
          $LL(j)$ = InitLiveline($l$, $z$)\\
          \\
          ParallelDFS() \{\\
          \quad\=\textbf{while} (not terminated) \{\\
          \>\quad\=node $n$=Pop()\\
          \>\>ProcessNode($n$)\\
          \>\>\textbf{if} (stack empty) \textbf{break}\\
          \>\>Probe()\\
          \>\>Distribute()\\
          \>\>Reject()\\
          \>\}\\
          \>Reject()\\
          \>Steal()\\
          \>Probe()\\
          \}\\
          ProcessNode(node $n$) \{\\
          \>\textbf{foreach} (child $c$ of $n$) Push($c$)\\
          \}\\
          Steal() \{\\
          \>\textbf{for} ($j=1$; $j<=w \wedge$ stack not empty; $j++$) \{\\
          \>\>Send(RandomId(), REQUEST)\\
          \>\>$steal\_replied$ = false\\
          \>\>\textbf{while} (true) \{\\
          \>\>\quad\=Probe()\\
          \>\>\>\textbf{if} ($steal\_replied$ == true) \textbf{break}\\
          \>\>\}\\
          \>\}\\
          \>\textbf{for} ($j=1$; $j<=z \wedge$ stack not empty; $j++$) \{\\
          \>\>\textbf{if} ($LL(j)$.activated == false) \{\\
          \>\>\>Send($LL(j)$, REQUEST)\\
          \>\>\}\\
          \>\}\\
          \}\\
          Probe() \{\\
          \>\textbf{while} (message received) \{\\
          \>\>for message (source, TYPE, payload)\\
          \>\>\textbf{switch} (message:TYPE) \{\\
          \>\>\>case REQUEST\\
          \>\>\>\quad\=\textbf{if} (stack empty) \{\\
          \>\>\>\>\quad\=Send(source, REJECT)\\
          \>\>\>\>\} \textbf{else} \{\\
          \>\>\>\>\>work = half of node stack\\
          \>\>\>\>\>Send(source, GIVE, work)\\
          \>\>\>\>\}\\
          \>\>\>case REJECT\\
          \>\>\>\>$steal\_replied$ = true\\
          \>\>\>case GIVE\\
          \>\>\>\>merge message:payload and local node stack\\
          \>\>\>\>LL($j$ for source).activated = false\\
          \>\>\>\>$steal\_replied$ = true\\
          \>\>\}\\
          \>\}\\
          \}
        \end{tabbing}
      \end{minipage}
    }
  \end{center}
  \caption{Parallel DFS with load balancing}
  \label{alg:parallelDFS}
\end{figure}

To reduce the number of communications,
Steal is called only if node stack is empty
and nodes are sent only if a request is received.

It is omitted from algorithm in the figure,
but if termination is detected,
finish message is broadcast and the algorithm exits the outer most while loop.
Termination detection is briefly described in
Section~\ref{sec:DTD}.

It became known recently that
the diameter of random graph is small~\cite{Shin:2011fd}.
Global Load Balancing (GLB) method was proposed in~\cite{Saraswat:2011cd}
which distributes workloads
following hypercube edges and random edges.
We are using this method for workload distribution.

GLB constructs a hypercube connecting all the processes,
which is called the {\it lifeline} graph.
When there are $P$ processes,
the lifeline with length $l$ should have
smallest possible dimension $z$ where $P \leq l^z$ holds.
Lifeline is constructed so that
$LL(j) \; (0 \leq j < z)$ is the process
id of the $j$-th lifeline neighbor.

GLB tries the random steal for $w$ times
and lifeline steal for $z$ times
until one of them succeeds.
We set $l=2$ (the hypercube has the highest possible dimensions) and
$w=1$ based on preliminary experiments and our past experience
(\cite{Ishii:2015kq}).



Requests are sent only if
the local stack becomes empty.
Only one REQUEST is sent at a time.
The algorithm waits until the request
is replied either by REJECT or GIVE.
When rejected, a request will be sent to
the next target.
If one steal phase finishes,
it will be in {\it idle} state
until it receives a GIVE message
from one of the processes connected by lifeline graph.

\subsection{Distributed termination detection}
\label{sec:DTD}

One of the problems which arises
for distributed algorithms is
Distributed Termination Detection (DTD).
In stack based parallel search,
workload can be created by any processes.
At first glance, it seems if stacks of all processes are empty,
the algorithm can terminate.
However that is not true because there can be a message
which is on the way from one process to another.

Mattern proposed several variations of
the {\it time algorithm} in~\cite{Mattern:1987iy}.
In time algorithm,
each process maintains a time-stamp
and a counter for recording
the difference of send/recv messages.

Messages for DTD are called {\it control} messages
and others are called {\it basic} messages.
When sending a basic message, the message counter is incremented
and on receiving a basic message, the counter is decremented.
All {\it basic} messages carry a time-stamp which is equal
to the time-stamp in the sender process.

The basic idea is to collect the message counter of all processes
and if the sum becomes zero, there will be no on-going messages.
However, it may detect false termination
if the same number of sends and receives are overlooked
because of the timing of gathering the message counter.


To prevent this case, each basic message carry
a time-stamp to check whether a message
had crossed the boundary of ``past'' and ``future''.
Past means the state before the collection
of the message counter and future means the state
after the collection.

Mattern proposed a {\it bounded clock-counter}
variant of the time algorithm.
It was originally proposed for a star topology
where one process sends control messages to all
other processes.
It could be easily modified for
a version using a spanning tree
and we have implemented
a version using a ternary tree.




\subsection{Closed itemset counter gather and broadcast}

As described in Section~\ref{subsec:support_increase},
the first phase of LAMP requires
to collect the sum of the number closed itemsets
and broadcast the updated $\lambda$ value.
It should be
frequently enough to avoid redundant computation
but at the same time it should not
disturb the main computation too much.

It is natural to implement such
gather and broadcast using a spanning tree.
In our case, Distribute Termination Detection
messages are sent using a spanning tree.
Therefor the closed itemset counter
is included in the payload of the DTD messages.
Fortunately, the delay of knowing the global value
only slows down the algorithm
and does not affect the correctness.









\subsection{Preprocess}

When starting search,
the root node will be pushed only to the
stack of the root process (which is proc. 0),
before the algorithm starts.
However, at the initial stage,
it is possible to start the search by
distributing the depth-1 children equally to
all processes.

If there are $P$ processes,
process with id $p_i$ only searches
item id $i$ if $i \mod P = p_i$.
Also, in the first phase,
we can count the closed itemsets after the preprocess phase
and we can start with $\lambda$ greater than 1,
to avoid frequent update of $\lambda$ in the beginning.



\subsection{Other Implementation details}

All code is written in c++
and parallelized using
the Message Passing Interface Library.

Our original target was
dense database with relatively small number of transactions.
Therefore we decided to exclude
database reduction technique from our implementation
and use the population count instruction
(counts the number of 1 on a register)
for counting the support.

In our algorithm,
one call to ProcessNode can take approximately
0.1 s at maximum,
and it would be too long for a time period
between Probe calls.
ProcessNode is modified so that
Probe and Distribute are called
approximately once in 1 millisecond.

\section{Experiments}
\label{sec:experiments}

\subsection{Setup of Experiments}

We applied our method to real datasets as shown in Table~\ref{tab:problems}. 
The HapMap and Alzheimer (Alz) data obtained from two
genome-wide association studies (GWASs),
which is widely conducting research to detect causal genes of disease
(Alzheimer study~\cite{Webster2009}, the HapMap project~\cite{HapMap2005}). 
A transaction and an item represent 
an individual and an existence of mutation, respectively, in both datasets.  
The positive and negative transactions were given as 
Alzheimer patient or not in Alz dataset. 
In HapMap dataset, 
Japanese are regarded as positive, 
and others are regarded as negative transactions. 

We generated two problems with different density from the same database
for testing our algorithm.
At first, we eliminated highly frequent items 
with the minor allele frequency (MAF), which is a widely used measure in GWAS analysis. 
When we used a high MAF threshold, 
the generated dataset contained high frequent items. 
Then, we calculated the existence of mutation for each individual 
based on dominant (dom) or recessive (rec) model. 
The results of dominant model has higher density than that of the recessive model. 

Additionally, 
to investigate effect on dataset 
containing small number of items and large number of transactions, 
we used human breast cancer transcriptome dataset (MCF7), 
which was analyzed in~\cite{Terada2013}. 

All experiments are performed on
TSUBAME supercomputer at Tokyo Institute of Technology.
Each compute node is equipped with
two Xeon X5670 processors (2.90GHz, 6 cores each),
and 52 GiB memory.
Nodes are connected with dual-rail QDR Infiniband network
with a total bandwidth of 80 Gbps.
We have used MVAPICH, an implementation of MPI library
and compiled the code using Intel compiler (ver. 14.0.2).

Performance of the parallel version
is averaged for at least 10 runs
because of the fluctuations caused by various reasons
such as network and/or compute node status.
Single thread performance is more stable
and averaged over at least four runs.

\subsection{Large scale performance}

\begin{figure*}[t]
  \centering
  \begin{tabular}{rrr}
    \multicolumn{1}{c}{HapMap dom. upper10} &
    \multicolumn{1}{c}{Alzheimer dom. upper05} &
    \multicolumn{1}{c}{Alzheimer rec. upper30} \\
    \includegraphics[scale=1.0]{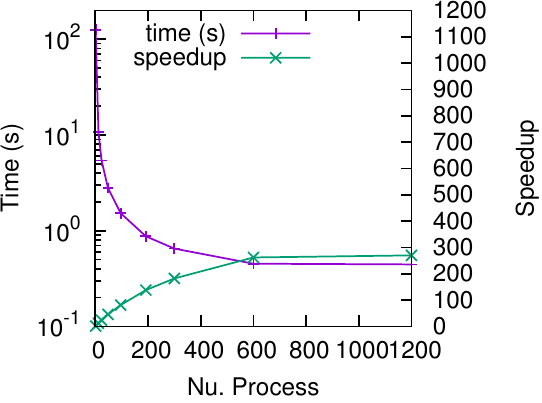} &
    \includegraphics[scale=1.0]{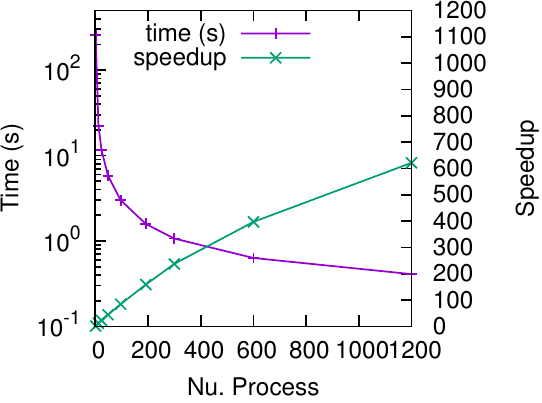} &
    \includegraphics[scale=1.0]{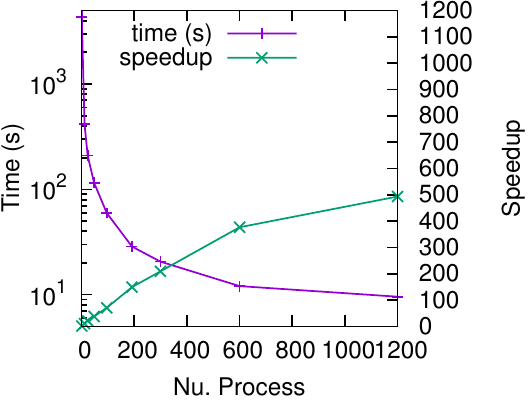} \\

    \multicolumn{1}{c}{HapMap dom. upper20} &
    \multicolumn{1}{c}{Alzheimer dom. upper10} &
    \multicolumn{1}{c}{MCF7 transcriptome} \\
    \includegraphics[scale=1.0]{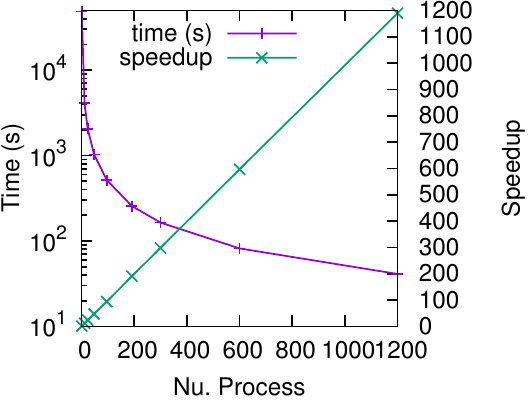} &
    \includegraphics[scale=1.0]{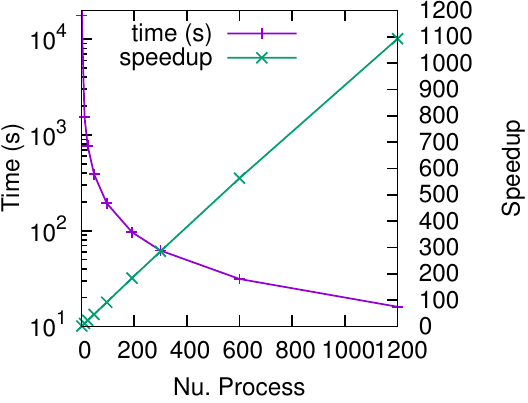} &
    \includegraphics[scale=1.0]{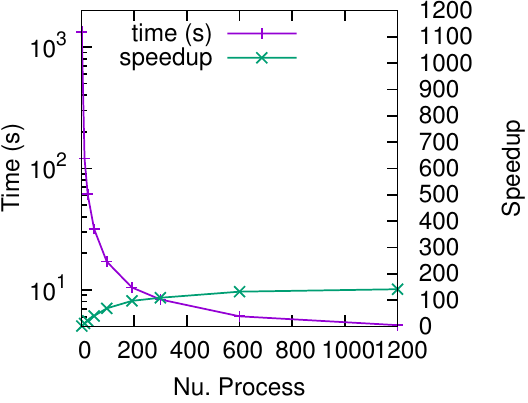} \\
  \end{tabular}
  \caption{Time and speedup}
  \label{fig:speedup}
\end{figure*}

The experimental results
using up-to 1,200 cores is shown in Fig.~\ref{fig:speedup}.
The speed was measured for
1, 12, 24, 48, 96, 192, 300, 600, and 1200 cores.
For two of the largest problems (Alz. dom. upper10 and HapMap dom. upper 20),
the speedup was almost linear to the number of processes
and this is very efficient as a parallel search algorithm.

Parallel algorithms tend to perform worse for
shorter execution time
because it is difficult to hide the overheads
given by communication and other operations due to parallelization.
However, for other smaller problems in genomics data,
our algorithm achieved an impressive performance.
Even for computational time shorter than 1.0 second
(measured on wall-clock time)
approximately 300 to 600-fold speedup was observed.

The breakdown of the total CPU time summed up for all processes
is given in Fig.~\ref{fig:breakdown}.
The time for 1 process, shown in the left-most bar of the figures,
are measured for our best single process algorithm
without any overhead needed for parallelization.

{\it Main} and {\it preprocess} are the computational time
required for the main computation for the search.
In ideal cases with linear speedup, the time indicated by the sum of
the gray and black bar ({main and preprocess)
should be equal to the 1 process bar.

The idle and probe categories
show the overhead of parallelization.
Probe part includes the time needed for send / receive of all message
and for splitting and merging the stack.
Idle part shows the time of waiting for replies
of steal requests or waiting for other processes to terminate.
The length of the probe and idle part does not largely differ for
genomics data.
It is clearly shown that for large problems,
the overhead is hidden by the longer computational time.

We haven't conducted experiments on slow network
such as Ethernet because
we didn't have such old cluster anymore.
However, we can estimate the effect of network delay.
The network delay only affects
a portion of the probe part
and if probe takes several more time,
the performance will be still good.

The last problem, MCF7,
is not our original target and the behavior is different from the
other five problems.
When using 600 or more cores,
the preprocess time (searching depth 1 nodes)
is taking longer than the rest of the computation.
This is because
there are only 397 items and it is smaller
than the number of processes
and all processes waits for other processes to finish
the preprocess.
This synchronization can be
removed but it is a part of the future work.

One advantage of our algorithm is that
no degradation of the speed was observed when increasing the number of cores.
For parallel search algorithms, the degradation is often observed
because of overhead.

\begin{figure*}[t]
  \centering
  \begin{tabular}{rrr}
    \multicolumn{1}{c}{HapMap dom. upper10} &
    \multicolumn{1}{c}{Alzheimer dom. upper05} &
    \multicolumn{1}{c}{Alzheimer rec. upper30} \\
    \includegraphics[scale=1.0]{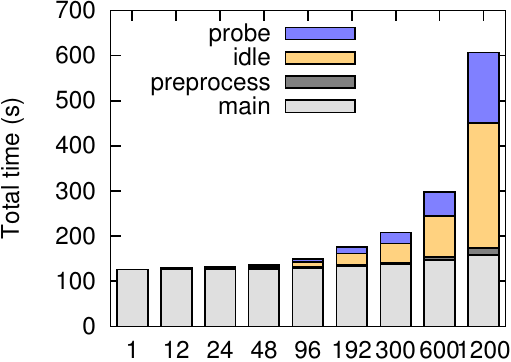} &
    \includegraphics[scale=1.0]{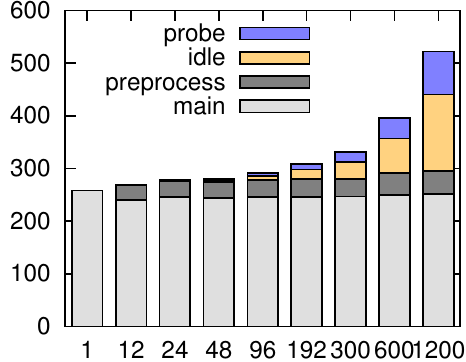} &
    \includegraphics[scale=1.0]{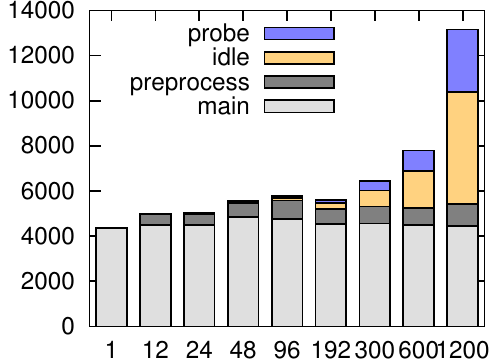} \\

    \multicolumn{1}{c}{HapMap dom. upper20} &
    \multicolumn{1}{c}{Alzheimer dom. upper10} &
    \multicolumn{1}{c}{MCF7 transcriptome} \\
    \includegraphics[scale=1.0]{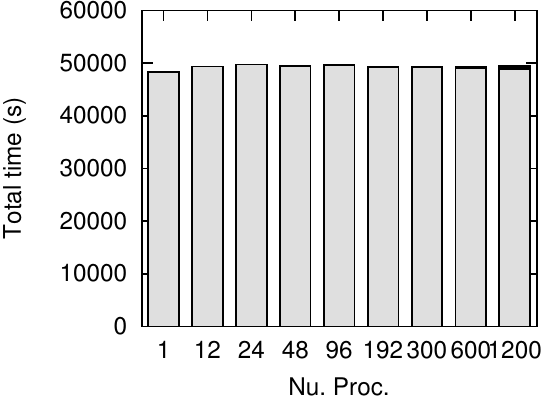} &
    \includegraphics[scale=1.0]{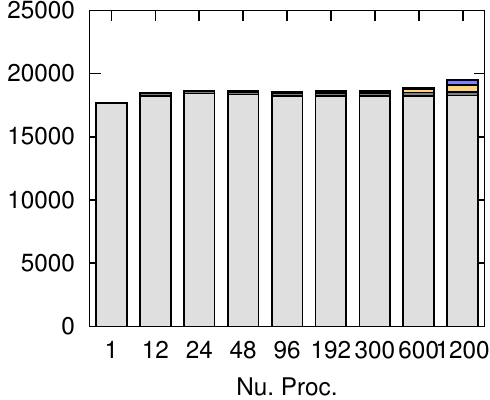} &
    \includegraphics[scale=1.0]{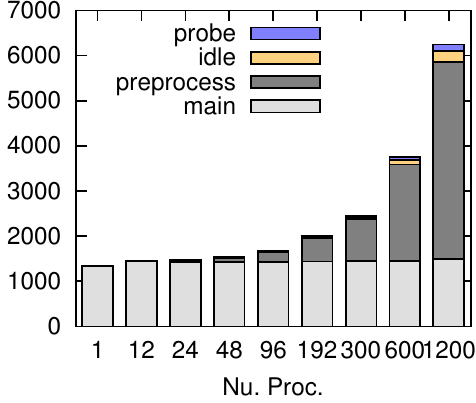} \\
  \end{tabular}
  \caption{Breakdown of total CPU time. Y-axis in seconds and X-axis shows nu. proc.}
  \label{fig:breakdown}
\end{figure*}

\begin{table*}
  \centering
  \caption{Problems.}
  \label{tab:problems}
  \begin{tabular}{lrrlrrrrrrr} \hline
    Name         & items   & trans. & density & $N_{pos}$ & $\lambda$ & nu. CS     &    $t_1$ & $t_{12}$ & $t_{1,200}$ \\ \hline
    HapMap dom. 10  & 11,253  & 697    & 1.02\% &   105 &             8 &     90,999 &      126 &   10.7  &   0.444 \\
    HapMap dom. 20  & 11,914  & 697    & 1.91\% &   105 &            11 & 47,835,176 &   48,285 &  4,108  &    41.1 \\
    Alz. dom.  5 & 44,052  & 364    & 5.40\% &   176 &            18 &     38,873 &      258 &   22.4  &    0.409 \\
    Alz. dom. 10 & 91,126  & 364    & 9.78\% &   176 &            23 &  1,113,223 &   17,646 &  1,535  &    16.0 \\
    Alz. rec. 30 & 250,120 & 364    & 2.90\% &   176 &            20 &    155,905 &    4,361 &    415  &    9.58 \\ \hline
    MCF7         & 397     & 12,773 & 2.94\% & 1,129 &             8 &  3,750,336 &    1,330 &    121  &    5.11 \\ \hline
  \end{tabular}
\end{table*}

\subsection{Performance on single computer}

The summary of the performance is shown in Table~\ref{tab:problems}.
Column $t_{12}$ on Table~\ref{tab:problems}
shows the performance on a single compute node
which has 12 CPU cores.

MPI library can be used a single compute node with small difficulty.
Communication is replaced with a memory copy which might
cause a small overhead but the results shows that
our algorithm achieved 10.5 to 11.8-fold speedup using 12 cores.
The results show the usefulness of our approach
for shared memory (e.g. single computer) parallelization.

\subsection{Limitations of Naive approach}

A simpler parallelization
can be done by just assigning part of the search space
to each process.
The performance of such naive algorithm
could be estimated by our algorithm
without any work steal.
(It still does broadcast of closed itemset number.)

Table~\ref{tab:vs_naive} shows the results
for single, 12, and 48 processes.
Naive approach shown in $n_{12}, n_{48}$ never overcome
our approach $t_{12}, t_{48}$.
For small problems it tends to perform better
which is because the search space is very shallow
and most part of the computation finishes within depth 1.
For large problems with deeper search spaces,
naive approach has a clear limit.

\begin{table*}
  \centering
  \caption{Comparison with Naive approach and LAMP2}
  \label{tab:vs_naive}
  \begin{tabular}{l|rD{.}{.}{1}D{.}{.}{2}|D{.}{.}{1}D{.}{.}{2}|D{.}{.}{1}D{.}{.}{2}D{.}{.}{3}|D{.}{.}{2}} \hline
    & \multicolumn{5}{c|}{vs. naive approach}
    & \multicolumn{4}{c}{vs. LAMP2 for 1st phase} \\
    Name         & \multicolumn{1}{c}{$t_1$} & \multicolumn{1}{c}{$t_{12}$} & \multicolumn{1}{c|}{$t_{48}$} &
    \multicolumn{1}{c}{$n_{12}$} & \multicolumn{1}{c|}{$n_{48}$} &
    \multicolumn{1}{c}{$t_1$} & \multicolumn{1}{c}{$t_{12}$} &
    \multicolumn{1}{c|}{$t_{1,200}$} & \multicolumn{1}{c}{$t_{LAMP2}$}
    \\ \hline
    HapMap dom. 10  &    126 &  10.7 & 2.79  & 13.7    & 7.26     &  58.7   &    4.91 &     0.218 & 2.69     \\
    HapMap dom. 20  &  48285 &  4108 & 1029  & 6559    & 3611     & 20973   &    1780 &      17.6 & 2083    \\
    Alz. dom.  5 &    258 &  22.4 & 5.80  & 24.1    & 9.90     & 131     & 11.3    & 0.211     & 78.6  \\
    Alz. dom. 10 &  17646 &  1535 & 387   & 3486    & 3480     & 9066    & 788     & 8.30      & 8582    \\
    Alz. rec. 30 &   4361 &   415 & 115   & 657     & 398      & 2239    & 209     & 4.54      & 1499   \\ \hline
    MCF7         &   1330 &   121 & 31.7  & 385     & 387      & 611     &    55.6 & 4.34      & 23.63 \\ \hline
  \end{tabular}
\end{table*}
  
\subsection{Comparison with LAMP2 (e.g. LCM)}

LAMP2 described in~\cite{Minato2014} uses
LCM ver. 5.3 as the base tool.
Since our code is tuned for large and high density database,
LAMP2 outperforms our approach if both are run on single core,
as shown in the right of Table~\ref{tab:vs_naive}.
But it is shown that for large problems, which is our main target,
the difference was small and
with 12 cores our result outperformed single process LAMP2.

However the results show that,
although it was not our main target, there is a large room for improvement
for sparse database with large number of transactions.

\subsection{Finding Significant Patterns}

As the final results,
we found statistically significant patterns
from the given database,
which are meaningful combinations of mutations in terms of genomics.
From HapMap dom. 20, we found statistically significant
itemsets having 8 items at maximum within less than 20 seconds,
which is far beyond the ability of brute force search.
The third phase took very short time (approx. 10 ms at most)
and the measurement is omitted from the paper.








\section{Related Work}
\label{sec:related_work}

\subsection{LAMP}


Multiple testing procedures 
for significant pattern mining have been proposed~\cite{Terada2013,Minato2014,Sugiyama2015}. 
LAMP~\cite{Terada2013} uses Bonferroni-like multiple testing procedures 
with Tarone's P-value bound strategy~\cite{Tarone1990} 
to improve the sensitivity of the correction through frequent itemset mining. 
The first LAMP version used a breadth-first search for finding the optimal value, 
whereas a depth-first search is known to be efficient, 
and hence Minato et al. proposed a fast algorithm for LAMP 
using the depth-first search~\cite{Minato2014}. 
However, more acceleration is required 
to analyze dense and large dataset. 
Our parallel strategy is applicable to 
detecting not only significant itemset 
but also significant subgraph mining~\cite{Sugiyama2015}. 




\subsection{Parallel search}


As already described, our work is based on
recent progress in smarter workload distribution
using hypercube with random edges \cite{Saraswat:2011cd}.
A similar approach was applied to Numerical Constraints Satisfaction
Progress (NCSP) and achieved approximately 500-fold speedup on 600 cores
\cite{Ishii:2015kq}.

There are other approaches for
parallel search algorithms which requires
hash table or priority queue, such as A* search or Iterative Deepening A*
(IDA*) search.
IDA* can be parallelized by distributing the hash table
based on a hash function as shown in TDS approach \cite{Romein:1999vy}.
A* search requires priority queue and seems
more difficult, but efficient performance
was achieved by Hash Distributed A* (HDA*) algorithm
\cite{Kishimoto:2013:ESS:2435476.2435960},
which is applied to planning problems.
Algorithms which does more pruning tend to be more difficult for
parallelization.
However, even for two player games where most of the branches are pruned,
successful parallel algorithms are reported,
for example, parallel Monte Carlo Tree Search algorithm \cite{Yoshizoe:2011uv}.

\subsection{Parallel Itemset Mining}




Speedup using up to 32 nodes (32 processors) is
reported for parallelization of the variants of the apriori algorithm
in \cite{agrawal1996parallel}.
Although part of the basic idea is common,
many of the enhancements are specific to apriori algorithm
and will not be needed for our LCM based solver.

Recent work on shared memory environment
include~\cite{Tatikonda:2009:MTD:1687627.1687706}
and~\cite{paraminer}.
These papers focus on a single computer and uses up-to 32 cores
and the reported speedup was efficient.
However, it is presumed that
directly applying these results to distributed memory
environment
requires re-designing of the algorithm as well as large implementation effort.



\section{Conclusions and Future Work}
\label{sec:conclusion}


Our parallel algorithm achieved,
260 to 1175-fold speedup using 1,200 cores.
Naturally, the scalability was better for more difficult problems
which should be the main targets of a parallel approach.
As a result, our approach is exceedingly promising.


The key difference between traditional algorithms
and massive parallel algorithm
is not whether memory is shared or distributed.
From the viewpoint of the algorithm,
the memory on different computer is just a slow and large memory.
Future algorithms have to deal with such environments.
We expect our algorithm continue to be efficient in the future
where an increased number of cores and memory is available.





\vspace{10pt}
\paragraph{Acknowledgments}

KY is supported by JST ERATO, Kakenhi 25700038, 15H02708.
AT is supported by JSPS Research Fellowships for Young Scientists.
KT is supported by JST CREST,
RIKEN PostK, NIMS MI2I, Kakenhi Nanostructure, Kakenhi 15H05711.





\bibliographystyle{IEEEtran}

\bibliography{bib_mplamp}

\begin{thebibliography}{10}
\providecommand{\url}[1]{#1}
\csname url@samestyle\endcsname
\providecommand{\newblock}{\relax}
\providecommand{\bibinfo}[2]{#2}
\providecommand{\BIBentrySTDinterwordspacing}{\spaceskip=0pt\relax}
\providecommand{\BIBentryALTinterwordstretchfactor}{4}
\providecommand{\BIBentryALTinterwordspacing}{\spaceskip=\fontdimen2\font plus
\BIBentryALTinterwordstretchfactor\fontdimen3\font minus
  \fontdimen4\font\relax}
\providecommand{\BIBforeignlanguage}[2]{{%
\expandafter\ifx\csname l@#1\endcsname\relax
\typeout{** WARNING: IEEEtran.bst: No hyphenation pattern has been}%
\typeout{** loaded for the language `#1'. Using the pattern for}%
\typeout{** the default language instead.}%
\else
\language=\csname l@#1\endcsname
\fi
#2}}
\providecommand{\BIBdecl}{\relax}
\BIBdecl

\bibitem{agrawal1996parallel}
R.~Agrawal and J.~C. Shafer, ``Parallel mining of association rules,''
  \emph{IEEE TKDE}, no.~6, pp. 962--969, 1996.

\bibitem{paraminer}
B.~Negrevergne, A.~Termier, M.-C. Rousset, and J.-F. M{\'e}haut,
  ``\BIBforeignlanguage{English}{Paraminer: a generic pattern mining algorithm
  for multi-core architectures},'' \emph{\BIBforeignlanguage{English}{Data Min.
  Knowl. Discov.}}, vol.~28, no.~3, pp. 593--633, 2014.

\bibitem{Tatikonda:2009:MTD:1687627.1687706}
S.~Tatikonda and S.~Parthasarathy, ``Mining tree-structured data on multicore
  systems,'' \emph{Proc. VLDB Endow.}, vol.~2, no.~1, pp. 694--705, 2009.

\bibitem{DBLP:conf/icdm/BuehrerPC06}
G.~Buehrer, S.~Parthasarathy, and Y.~Chen, ``Adaptive parallel graph mining for
  {CMP} architectures,'' \emph{Proc. {ICDM} 2006}, pp. 97--106, 2006.

\bibitem{DBLP:conf/ieeehpcs/NegrevergneTMU10}
B.~N{\'{e}}grevergne, A.~Termier, J.~M{\'{e}}haut, and T.~Uno, ``Discovering
  closed frequent itemsets on multicore: Parallelizing computations and
  optimizing memory accesses,'' \emph{Proc. {HPCS} 2010}, pp. 521--528, 2010.

\bibitem{corley2010intel}
A.-M. Corley, ``Intel lifts the hood on its ``single-chip cloud computer'',''
  \emph{IEEE Spectrum Online (9 feb 2010)}, 2010.

\bibitem{uno2004lcm}
T.~Uno, M.~Kiyomi, and H.~Arimura, ``{LCM} ver. 2: Efficient mining algorithms
  for frequent/closed/maximal itemsets,'' \emph{Proc. {IEEE} {ICDM'04} Workshop
  {FIMI'04}}, vol. 126, 2004.

\bibitem{Saraswat:2011cd}
V.~A. Saraswat, P.~Kambadur, S.~Kodali, D.~Grove, and S.~Krishnamoorthy,
  ``{Lifeline-based global load balancing},'' \emph{Proc. PPoPP '11}, 2011.

\bibitem{maher2008personal}
B.~Maher, ``Personal genomes: The case of the missing heritability,''
  \emph{Nature}, vol. 456, no. 7218, pp. 18--21, 2008.

\bibitem{onkamo2006survey}
P.~Onkamo and H.~Toivonen, ``A survey of data mining methods for linkage
  disequilibrium mapping,'' \emph{Hum Genomics}, vol.~2, no.~5, pp. 336--340,
  2006.

\bibitem{Terada2013}
A.~Terada, M.~Okada-Hatakeyama, K.~Tsuda, and J.~Sese, ``{Statistical
  significance of combinatorial regulations},'' \emph{Proc Natl Acad Sci U S
  A.}, vol. 110, no.~32, pp. 12\,996--13\,001, 2013.

\bibitem{Minato2014}
S.~Minato, T.~Uno, K.~Tsuda, A.~Terada, and J.~Sese, ``{A Fast Method of
  Statistical Assessment for Combinatorial Hypotheses Based on Frequent Itemset
  Enumeration},'' \emph{ECML/PKDD 2014}, vol. 8725, pp. 422--436, 2014.

\bibitem{Kishimoto:2013:ESS:2435476.2435960}
A.~Kishimoto, A.~Fukunaga, and A.~Botea, ``Evaluation of a simple, scalable,
  parallel best-first search strategy,'' \emph{Artif. Intell.}, vol. 195, pp.
  222--248, 2013.

\bibitem{Romein:1999vy}
J.~W. Romein, A.~Plaat, H.~E. Bal, and J.~Schaeffer, ``{Transposition Table
  Driven Work Scheduling in Distributed Search.}'' \emph{AAAI/IAAI}, pp.
  725--731, 1999.

\bibitem{Dudoit2007}
S.~Dudoit and M.~J. {Van Der Laan}, \emph{{Multiple Testing Procedures and
  Applications to Genomics}}, 2007.

\bibitem{Tarone1990}
R.~E. Tarone, ``{A modified Bonferroni method for discrete data},''
  \emph{Biometrics}, vol.~46, no.~2, pp. 515--522, 1990.

\bibitem{Shin:2011fd}
J.-Y. Shin, B.~Wong, and E.~G. Sirer, ``{Small-world datacenters},''
  \emph{Proc. SOCC '11}, 2011.

\bibitem{Ishii:2015kq}
D.~Ishii, K.~Yoshizoe, and T.~Suzumura, ``{Scalable parallel numerical
  constraint solver using global load balancing},'' \emph{Proc. X10 2015}, pp.
  33--38, 2015.

\bibitem{Mattern:1987iy}
F.~Mattern, ``{Algorithms for distributed termination detection},''
  \emph{Distributed Computing}, vol.~2, no.~3, pp. 161--175, 1987.

\bibitem{Webster2009}
J.~A. Webster, J.~R. Gibbs, J.~Clarke \emph{et~al.}, ``{Genetic control of
  human brain transcript expression in Alzheimer disease.}'' \emph{Am J Hum
  Genet.}, vol.~84, no.~4, pp. 445--58, 2009.

\bibitem{HapMap2005}
{The International HapMap Consortium}, ``{A haplotype map of the human
  genome.}'' \emph{Nature}, vol. 437, no. 7063, pp. 1299--320, 2005.

\bibitem{Sugiyama2015}
M.~Sugiyama, F.~Llinares-L{\'o}pez, N.~Kasenburg, and K.~M. Borgwardt,
  ``Significant subgraph mining with multiple testing correction,'' \emph{SIAM
  SDM15}, pp. 37--45, 2015.

\bibitem{Yoshizoe:2011uv}
K.~Yoshizoe, A.~Kishimoto, T.~Kaneko, H.~Yoshimoto, and Y.~Ishikawa,
  ``{Scalable Distributed Monte-Carlo Tree Search},'' \emph{SoCS'11}, pp.
  180--187, 2011.

\end{thebibliography}

\end{document}